\documentclass[a4paper,12pt]{extarticle}
\usepackage{geometry}
\usepackage[english]{babel}
\usepackage{amsmath,amssymb,amsfonts,amsthm}
\usepackage[mathscr]{eucal}
\usepackage[all]{xy}
\usepackage{hyperref}
\usepackage{setspace}
\usepackage{upgreek}
\usepackage{hyperref}
\usepackage{cite}
\usepackage{graphicx}
\usepackage{authblk}

\usepackage{times}
\usepackage{color}

\usepackage{indentfirst}

\allowdisplaybreaks
\title{Generalized Maxwell-Jüttner distribution \\
for rotating spinning particle gas}
\author{Dmitry S. Kaparulin}
\author{Nikita N. Levin}
\affil{Tomsk State University, Tomsk 634050, Russia\\
e-mail: dsc@phys.tsu.ru, levinnn@phys.tsu.ru}

\begin{document}
\maketitle
\begin{abstract}
    We consider a statistical mechanics and thermodynamics of a rotating ideal gas of classical relativistic particles with nonzero mass and spin. Applying the Gibbs theory of canonical ensembles for a system rotating with constant angular velocity, we obtain the one-particle distribution function by positions, momenta, and spin variables. By computing the partition function, we obtain various thermodynamic quantities of slowly rotating gas. Both the statistical and thermodynamic approaches demonstrate the polarization of spinning degree of freedom, with the majority of spins being directed along the angular velocity vector. This confirms the presence of chiral effects in the system. 
\end{abstract}

\section{Introduction}
The statistical mechanics of (non-)ideal gases has been studied for almost two centuries. The well-known results in the field include the Maxwell and Maxwell-Boltzmann distributions for momenta and positions of the particles \cite{LL}. They are applicable in the case of not too high temperatures when the kinetic energy of the particles is small with respect to their rest mass. For higher temperatures, the Maxwell-Boltzmann distribution is replaced by the Maxwell-Jüttner distribution \cite{MJ}. The Maxwell-Boltzmann distribution for rotating gas has been developed by Maxwell \cite{JCM-1878}, while the canonical distribution function for the rotating classical system has been developed by Gibbs \cite{JWG-1902}. For recent reconsideration of the problem of rotating Boltzmann relativistic gas, we cite \cite{second reviewer ed2, AB-2015}. These works have one common feature: they consider a gas particle as a point object in space-time without any internal degrees of freedom. In such system, the main effect of rotation is the change of the particle concentration due to influence of centrifugal force. 

In the systems of spinning particles, the interaction of spin angular momentum and macroscopic rotation causes a special class of phenomena, known as the chiral effects. First time they have been observed in 1915 \cite{Barnett, EdH}. Now, we know that the chiral phenomena can influence almost all macroscopic characteristics of rotating matter, including internal energy, entropy, and heat capacity. At the microscopic level, the rotation implies the non-homogeneous distribution of directions of spins, with majority angular momenta being directed along the axis of rotation \cite{Becc}. A recent review on the subject can be found in \cite{Fukushima}. The current studies of chiral phenomena mostly use the quantum treatment of spin withing the formalism of quantum statistics and/or quantum field theory at the finite temperature. There is a vast literature on the subject \cite{Vilenkin, Fukushima1, Chernodub, Chernodub1,  Xu, ed1, ed2, ed3, ed4, ed5, ed6, ed7, ed8, ed9, ed10}. The studies include the relativistic hydrodynamics of liquids with spin, which is important in the context of nuclear collisions description \cite{She,Florkowski,Bhadury,Florkowski1}, and in astrophysics \cite{Sterg}. Significant progress is reached in formalism of chiral kinetic theory \cite{ckt1, ckt2}. For a review of studies of spin-polarized fluids, we cite \cite{second reviewer ed1}. 

The classical spin theory provides an interesting alternative to quantum studies of chiral effects. It has been developing since the works of Frenkel \cite{Fren}, for a modern review see \cite{Fryd} and \cite{Deriglazov}. In the article \cite{Universal}, it has been shown that the (charged) spinning particle of arbitrary spin admits consistent propagation at general electromagnetic or gravitational background. This seems to be an important advantage of the quasi-classical formalism, as the constriction of relativistic higher-spin equations on general backgrounds faces to obstructions already at the classical level \cite{Kaparulin, Cortese, Boulanger}. In the article \cite{DP-2014F}, the equations of motion of spinning particle in uniform magnetic field has been solved. It has been shown that the spin influences the cyclotron frequency, and also cause the new effect, the fluctuations of trajectory being termed "magnetic zitterbewegung". In the article \cite{KS-2023}, the Coulomb problem for the weakly relativistic classical spinning particle has been addressed. It has been shown, that the quasi-classical Borh-Sommerfeld quantization formula reproduces the energy levels of hydrogen atom with the precision up to the fifth order in the fine structure constant in the large quantum number limit - better than relativistic quantum mechanics can do.  

In the current article, we address the problem of construction of statistical mechanics of relativistic spinning particles using the quasi-classical description of spin. A similar problem for non-relativistic particles has been solved \cite{BKN-2022}. Our study has several motivations. The possibility of application of the relativistic theory to systems with external fields extends the class of macroscopic systems that admit description within the formalism of statistical mechanics. The small value of chiral effects suggests that the quasi-classical computations may provide valuable information to study properties of rotating systems. Another significant motivation relies on a technical argument. The quasi-classical computations are typically simple, and they give good chances for obtaining final answers in explicit form, being better suited for consequent analysis. The quasi-classical theory also allows construction of angular distributions by directions of spin, which has no analog in quantum theory. This angular distribution provides a simple visualization of chiral effects. Finally, the classical distribution function admits a simple taking of the ultra-relativistic limit, making it possible to study ultra-hot gases. 
 
In the text below, we address three particular questions concerning the rotation relativistic gas of spinning particles: construction of distribution function by momenta, positions and directions of spin, computation of partition function and thermodynamic potential of macroscopic system, and comparison of our conclusions with previous studies. By solving the first problem, we derive the generalized Maxwell-Jüttner distribution for a rotating gas of spinning particles. We find out that the distribution function depends on spin, so the rotation affects the state of gas. This confirms the presence of chiral effects. We also note that, for a relativistic particle, the spinning and translational degrees of freedom do not decouple. This demonstrates a some kind of spin-orbital interaction, which does not have analogue in the non-relativistic gas. Solution of the second problem leads us to the thermodynamic potential of spinning gas in a rotating cylinder. Using it, we find chemical potential, entropy, and total angular momentum of the gas. The obtained formulas are significantly simplified in the ultra-relativistic case. In the concluding part of the article, we compare the results of our study with the previous works \cite{Becc, second reviewer ed2}, and \cite{BKN-2022}.

The article is organized as follows. In the next section, we make an exposition of a spinning particle concept. We mostly focus on the construction of appropriate invariant measure on the phase-space, which is required by the classical statistical mechanics. In Section 3, we apply the Gibbs distribution for rotating systems to obtain a one-particle distribution function in different coordinate systems. In Section 4, we consider the angular distribution function, that characterizes the behavior of spinning degree of freedom. In Section 5, we consider a thermodynamics of rotating gas and compute the average value of spin angular momentum. In Section 6 we compare obtained results with results of Ref. \cite{Becc}. Conclusion summarizes the results. 

\section{Microscopic description}

In this section, we recall basic facts about the description of dynamics of classical spinning particle that are required for construction of statistical mechanics. We consider a model of massive irreducible relativistic spinning particle with the mass $m$ and spin $s$ with vector parameterization of spin states space, previously proposed in \cite{Deriglazov}. The model is connected to early theory, where a spinor parameterization of space of spin states is used \cite{Universal}. For the exposition of spinning particle concept, we cite \cite{Fryd}. As the application to statistical mechanics is in question, we focus on two particular issues: the construction of phase-space of the particle (including invariant measure on it), and expression of the energy-momentum in terms of phase-space coordinates. The problem is non-trivial because all the current spinning particle theories have gauge freedom, complicating the construction of physical observables and  invariant measure. To solve this problem, we explicitly solve all the constraints in the theory. The reduced phase-space is shown to be a fiber bundle, $\mathbb{R}^6\times\mathbb{S}^2$ as it is expected for a massive spinning particle. The coordinates on the phase-space include the six global coordinates on $\mathbb{R}^6$ and two angular variables on $\mathbb{S}^2$. Then, we express angular momentum and total angular momentum in terms of phase-space coordinates. The latter problem is quite technical, but it is necessary for making explicit computations. 

We consider a spinning particle as a point object in Minkowski space. The space-time coordinates of the particle are denoted by $r^\mu,\mu=0,1,2,3$. The canonically conjugated momenta are denoted by $p^\mu$. To describe the spin, we introduce a vector variable $\omega^\mu$, and conjugated momenta $\pi^\mu$. We assume that the tensor indices are raised and lowered by Minkowski metric $g_{\mu\nu}$. We use mostly positive signature for the metric throughout the paper. The round brackets denote the scalar product of two space vectors with respect to the metric. 
The phase-space variables are subjected to constraints,
\begin{equation}\label{constr}\begin{array}{c}\displaystyle
    \Theta_p=(p,p)+m^2c^2\,,\quad \Theta_\omega=(\omega,\omega)-a^2\,,\quad\Theta_\pi=(\pi,\pi)-b^2\,,\quad \Theta_{p\omega}=(p,\omega)\\[3mm]\Theta_{p\pi}=(p,\pi)\,,\quad\Theta_{\omega\pi}=(\omega,\pi)\,.
\end{array}\end{equation}
The quantity $m$ denotes the mass of the particle, and $c$ is the speed of light. The parameters $a$, $b$ determine the particle spin by the rule $a^2b^2=\hbar^2 s(s+1)$. In what follows, the norm of the spin vector (in the Plank units) is denoted by symbol 
\begin{equation}
    \Sigma=\sqrt{s(s+1)}\,.   
\end{equation}
The quantity $S$ can be considered as the "classical" spin of the particle. In contrast to $s$, it is an irrational number. We do not fix the particular values of $a$ and $b$ because only the product $ab$ is relevant in the context of statistical description. We chose the constrained Hamiltonian action derived in \cite{Deriglazov},
\begin{equation}\label{S-Ham}
    S=\int\bigg\{(p,\dot{r})+(\pi,\dot{\omega})-\lambda_p\Theta_p-\lambda_\omega\Theta_\omega-\lambda_\pi\Theta_\pi-\lambda_{p\omega}\Theta_{p\omega}-\lambda_{p\pi}\Theta_{p\pi}-\lambda_{\omega\pi}\Theta_{\omega\pi}\bigg\}d\tau\,.
\end{equation}
Here, the dynamical variables are particle position in space-time $r^\mu$, linear momentum $p^\mu$, spinning sector variables $\omega^\mu$ and $\pi^\mu$, and six Lagrange multipliers $\lambda_p,\ldots,\lambda_{\omega\pi}$. The world index runs over the values $\mu=0,1,2,3$. All the dynamical variables are functions of the proper time $\tau$. The dot denotes the derivative by $\tau$.

The action functional corresponds to the irreducible spinning particle. This means that physical observables of the model are a function of momentum and total angular momentum. The momentum is given by the quantity $p^\mu$, while the total angular momentum reads
\begin{equation}\label{j-tot}
    j{}^{\mu\nu}=r{}^\mu p{}^\nu-r{}^\nu p{}^\mu+\omega{}^\mu \pi{}^\nu -\omega{}^\nu \pi{}^\mu\,,\qquad \mu,\nu=0,1,2,3\,.
\end{equation}
By construction, the total angular momentum is a sum of orbital angular momentum and spin angular momentum (Frenkel spin angular momentum in the terminology \cite{Deriglazov}). The latter is determined by the formula,
\begin{equation}\label{s-def}
    s{}^{\mu\nu}=\omega{}^\mu \pi{}^\nu -\omega{}^\nu \pi{}^\mu\,.
\end{equation}
The vector of spin angular momentum is traverse to momenta and normalized,
\begin{equation}\label{pS-SS}
    s{}_{\mu\nu}p^\nu=0\,,\qquad s{}_{\mu\nu}s{}^{\mu\nu}=4a^2b^2=4\hbar^2s(s+1)\,.
\end{equation}
Relations (\ref{pS-SS}) imply that the state of spinning degree of freedom is characterized by two variables, determining the direction of spin vector. The dimension of physical phase-space of a massive spinning particle is eight. The other variables are spacial positions of the particle, and its spacial linear momentum. 

In the context of statistical mechanics, the most critical step is the constriction of the invariant measure on the space. The problem is non-trivial due to presence of constraints and gauge symmetries in the model (\ref{S-Ham}). For our purposes, it is convenient to solve all the constraints, and introduce unconstrained coordinates on the physical phase-space, being equivalent to $\mathbb{R}^6\times\mathbb{S}^2$. In this case, the invariant norm will be the product of invariant norms on its factors. It is convenient to use $1+3$ decomposition for the generalized coordinates and momenta. By definition, we put
\begin{equation}\label{rp-13}
    r^\mu=(ct,\boldsymbol{r})\,,\qquad p^\mu=mc\gamma(1,\boldsymbol{\beta})\,,\qquad \omega^\mu=(\omega^0,\boldsymbol{\omega})\,,\qquad \pi^\mu=(\pi^0,\boldsymbol{\pi})\,.
\end{equation}
Here, the bold symbols represent the spacial coordinates of the corresponding vectors. The dimensionless quantities $\boldsymbol{\beta}$ and $\gamma$ denote standard relativistic factors, being associated with the motion of the particle. The norm of the vector $\boldsymbol{\beta}$ with respect to the Euclidean metric is denoted by $\beta$, $\beta^2=(\boldsymbol{\beta},\boldsymbol{\beta})$. The round brackets with enclosed Euclidean vectors denote Euclidean scalar product. The quantities $\beta$ and $\gamma$ are connected by the rule
\begin{equation}\label{bg-fac}
    \gamma=\frac{1}{\sqrt{1-\beta^2}}\qquad\Leftrightarrow\qquad\beta^2=1-\frac{1}{\gamma^2}\,.
\end{equation}
Our further strategy is to express the all time coordinates of spinning sector variables via their spacial components, and then, to find and appropriate parametric solution for Euclidean space vectors in terms of angular variables. We also use special notation for the spacial coordinates. By definition, we put $\boldsymbol{x}=(x,y,z)\,.$
The coordinates of Euclidean space vector are labelled by $x,y,z$ subscripts. For example, for a relativistic factor $\boldsymbol{\beta}$, we have the representation $\boldsymbol{\beta}=(\beta_x,\beta_y,\beta_z)$.

In the $1+3$ decomposition, the constraints (\ref{constr}) for spinning sector variables take the following form:
\begin{equation}\label{constr-1}
    \omega^0-(\boldsymbol{\beta},\boldsymbol{\omega})=0\,,\qquad\pi^0-(\boldsymbol{\beta},\boldsymbol{\omega})=0\,.
\end{equation}
\begin{equation}\label{constr-2}
    (\boldsymbol{\omega},\boldsymbol{\omega})-\omega^0\omega^0=a^2\,,\qquad (\boldsymbol{\pi},\boldsymbol{\pi})-\pi^0\pi^0=b^2\,,\qquad (\boldsymbol{\omega},\boldsymbol{\pi})-\omega^0\pi^0=0\,.
\end{equation}
The system includes five equations for eight unknown components of two four-vectors $\omega$, $\pi$. Relations (\ref{constr-1}) express time components $\omega^0$, $\pi^0$ in terms of $\boldsymbol{\omega}$, $\boldsymbol{\pi}$, and $\boldsymbol{\beta}$. Once $\omega^0$, $\pi^0$ are expressed in terms of $\boldsymbol{\omega}$, $\boldsymbol{\pi}$, the relations (\ref{constr-2}) determine a system of three equations for a pair of Euclidean vectors $\boldsymbol{\omega}$, $\boldsymbol{\pi}$. The solution to the system (\ref{constr-1}), (\ref{constr-2}) with respect to the unknowns $\boldsymbol{\omega}$, $\boldsymbol{\pi}$ reads
\begin{equation}\label{pw-sol}
    \boldsymbol{\omega}=a\bigg(\boldsymbol{a}+\frac{\gamma^2}{\gamma+1}\boldsymbol{\beta}(\boldsymbol{\beta},\boldsymbol{a})\bigg)\,,\qquad \boldsymbol{\pi}=b\bigg(\boldsymbol{b}+\frac{\gamma^2}{\gamma+1}\boldsymbol{\beta}(\boldsymbol{\beta},\boldsymbol{b})\bigg)\,.
\end{equation}
The solution involves a pair of normalized and orthogonal Euclidean space vectors $\boldsymbol{a}$, $\boldsymbol{b}$ such that
\begin{equation}
    (\boldsymbol{a},\boldsymbol{a})=1,\qquad (\boldsymbol{b},\boldsymbol{b})=1\,,\qquad (\boldsymbol{a},\boldsymbol{b})=0\,.
\end{equation}
The last vectors can be considered as the first and second basis vector of certain right-handed orthogonal frame in Euclidean space. The frame is unique because the third vector is given by the vector product $\boldsymbol{c}=[\boldsymbol{a},\boldsymbol{b}]$\,. The set of right-handed frames in three-dimensional Euclidean space is parameterized by the three Euler angles, which we denote $\alpha^\circ,\beta^\circ, \gamma^\circ$. Here, the circle means that the angular variable is involved into expression. The solution with respect to the vectors $\boldsymbol{a}$, $\boldsymbol{b}$, and $\boldsymbol{c}$ have the following form:
\begin{equation}\label{a-vec}
    \begin{array}{c}
        \boldsymbol{a}(\alpha^\circ, \beta^\circ, \gamma^\circ) =
        \left(
        \begin{array}{c}
        \phantom{+}\cos\alpha^\circ \cos\beta^\circ \cos\gamma^\circ - \sin\alpha^\circ \sin\gamma^\circ
        \\
       \phantom{+}\sin\alpha^\circ \cos\beta^\circ cos\gamma^\circ + \cos\alpha^\circ \sin\gamma^\circ
        \\
        -\sin\beta^\circ \cos\gamma^\circ
        \end{array}
        \right)\,;
    \end{array}
\end{equation}
\begin{equation}\label{b-vec}
    \begin{array}{c}
        \boldsymbol{b}(\alpha^\circ, \beta^\circ, \gamma^\circ) =
        \left(
        \begin{array}{c}
        -\cos\alpha^\circ \cos\beta^\circ \sin\gamma^\circ - \sin\alpha^\circ \cos\gamma^\circ
        \\
        -\sin\alpha^\circ \cos\beta^\circ \sin\gamma^\circ + \cos\alpha^\circ \cos\gamma^\circ
        \\
        \sin\beta^\circ \sin\gamma^\circ
        \end{array}
        \right)\,;
    \end{array}
\end{equation}
\begin{equation}\label{c-vec}
    \begin{array}{c}
        \boldsymbol{c}(\alpha^\circ, \beta^\circ, \gamma^\circ) =
        \left(
        \begin{array}{c}
        \cos\alpha^\circ \sin\beta^\circ
        \\
        \sin\alpha^\circ \sin\beta^\circ
        \\
        \cos\beta^\circ
        \end{array}
        \right)\,.
    \end{array}
\end{equation}
Here, the Euler angles $\alpha^\circ,\beta^\circ, \gamma^\circ$ run over the set 
\begin{equation}
    0\leq\alpha,\gamma\leq 2\pi\,,\qquad 0\leq\beta\leq\pi\,.
\end{equation}
The representation (\ref{pw-sol}), (\ref{a-vec}), (\ref{b-vec}), (\ref{c-vec}) for the vectors $\omega$, $\pi$ have a clear geometrical origin. In the rest system of the particle, $p^\mu=(mc,\boldsymbol{0})$, the four-vectors $\omega$, $\pi$ have zero time components. Then, the equations (\ref{constr-2}) ensure that their spacial components $\boldsymbol{\omega}$, $\boldsymbol{\pi}$ are orthogonal and normalized. Then $\boldsymbol{a}$, $\boldsymbol{b}$ denote unit vectors $\boldsymbol{\omega}/a$, $\boldsymbol{\pi}/b$. Solution is obtained (\ref{pw-sol}), (\ref{a-vec}), (\ref{b-vec}), (\ref{c-vec}) by the Lorentz transformation from the rest system to the system with the general value of momentum $p$.

In the context of the current study, it is important to
express the vector of spin angular momentum in terms of angular variables $\alpha^\circ$, $\beta^\circ$, $\gamma^\circ$. We define $1+3$ decomposition for the tensor of spin 
(\ref{s-def}) by the rule $s^{\mu\nu}=(\boldsymbol{l},\boldsymbol{s})$. 
The Euclidean space vector $\boldsymbol{l}$ with the coordinates $l^i=s^{0i}$ denotes the space-time components of the spin tensor. The Euclidean space vector $\boldsymbol{s}$ with the coordinates $s^i$ stands for the space components of the spin tensor. The quantities $s^{i}$ are determined by the rule 
\begin{equation}                    s^{ij}=\varepsilon^{ijk}s_k\qquad\Leftrightarrow\qquad s_{i}=\frac{1}{2}\varepsilon_{ijk}s^{jk}\,.
\end{equation}
 Substituting (\ref{pw-sol}), (\ref{a-vec}), (\ref{b-vec}), (\ref{c-vec}) into the definition of angular momentum (\ref{s-def}), we obtain the following expressions for $\boldsymbol{l}$, $\boldsymbol{s}$:
\begin{equation}\label{ls-vec}
    \boldsymbol{l} =-\hbar\Sigma\gamma [\boldsymbol{\beta}\,,\boldsymbol{c}]\,,\quad
    \boldsymbol{s} = \hbar\Sigma\bigg(\boldsymbol{c} - \frac{\gamma^2}{\gamma+1}[\boldsymbol{\beta}\,,[\boldsymbol{\beta}\,,\boldsymbol{c}]]\bigg)\,.
\end{equation}
Here, $\hbar$ is the Plank constant, $s$ (without indices) is the particle spin, and $\boldsymbol{c}$ is the vector (\ref{c-vec}); $\boldsymbol{\beta}$ and $\gamma$ are relativistic factors (\ref{rp-13}), (\ref{bg-fac}). Formulas (\ref{ls-vec}) allow us to parameterize the state of spinning degree of freedom by two angular variables $\alpha^\circ$, $\beta^\circ$. The other Euler angle $\gamma^\circ$ drops out of the final expression. This is not surprising because the rotations of the vectors $\boldsymbol{\omega}$, $\boldsymbol{\pi}$ is gauge freedom of the model. Combining (\ref{rp-13}), (\ref{c-vec}), and (\ref{ls-vec}), we express the particle energy, momentum, and total angular momentum in terms of eight independent variables: three spacial coordinates $\boldsymbol{x}=(x,y,z)$, three spacial components of the momenta $\boldsymbol{p}=(p_x,p_y,p_z)$, and two angles $\alpha^\circ$, $\beta^\circ$, being the coordinates on the space of particle states. The invariant norm on the phase-space $\mathbb{R}^6\times\mathbb{S}^2$ is given by the product of the norm on the factors. In the mentioned above coordinates it has an obvious form,
\begin{equation}\label{dg}
    \mathrm{d}\Gamma=\frac{(2s+1) \mathrm{d}\boldsymbol{x}\mathrm{d}\boldsymbol{p}\sin\beta^\circ \mathrm{d}\alpha^\circ \mathrm{d}\beta^\circ}{4\pi(2\pi\hbar)^3}
\end{equation}
We chose an overall factor from the requirement of relationship with quantum statistics. The quotient $(2\pi\hbar)^3$ represents the phase space volume occupied by one translational state. The factor $(2s+1)/4\pi$ accounts that the particle with spin $s$ has $(2s+1)$ possible states. We note the that the normalizing factor is a natural ambiguity in the classical statistics because the classical volume element $\mathrm{d}\boldsymbol{x}\mathrm{d}\boldsymbol{p}$ is a dimensional quantity. With the formula (\ref{dg}), we finalize the description of spinning particle at the microscopic level. In the following text, we begin dealing with the construction of statistical mechanics. 

\section{One-particle distribution function}

In this section, we address the question of construction of one-particle distribution function for a rotating relativistic ideal gas of spinning particles. 

Our starting point is the canonical Gibbs distribution for a rotating relativistic system \cite{LL},
\begin{equation}\label{f-pj}
    f_0(p,j)=\frac{1}{z_0}\exp\bigg(u^\mu 
 p_\mu+\frac{1}{2}w^{\mu\nu}j_{\mu\nu}\bigg).
\end{equation}
Here, the quantities $p_\mu$ and $j_{\mu\nu}$ represent the total momentum and total angular momentum of the system, and $u^\mu$, $w^{\mu\nu}$ are parameters of the distribution. The quantity $z$ is a normalizing factor. Throughout the article, we assume that $f$ is normalized by 1. The normalization is determined with respect to the canonical measure on the phase space. 

The structure of distribution function (\ref{f-pj}) unambiguously determined by the fact that $p_\mu$ and $j_{\mu\nu}$ are additive integrals of motion. The vector parameters $u^\mu(u^0,\boldsymbol{u})$ determine the thermodynamic temperature of the system (the parameter $u^0$), and the velocity of the motion of the mass center (the parameters $u^i$). After appropriate normalization, the tensor parameters determine the spacial acceleration (the quantities $w^{0i}$), and the spacial angular velocity (the quantities $w^{ij}$). In the current study, we consider a system having a center of mass at rest, so $u^i=0$ and $w^{0i}=0$. For the remaining parameters, we introduce a special pasteurization, $u^0=c/kT$ and $w^{ij}=\epsilon_{ijm}\Omega_m/kT$, with 
$k$ being the Boltzmann constant. The new scalar and vector parameters $T$ and $\boldsymbol{\Omega}$ are identified with the thermodynamic temperature of the system, and its angular velocity of macroscopic rotation. The rotation velocity at the point with the position $\boldsymbol{x}$ reads $[\boldsymbol{\Omega},\boldsymbol{x}]$. To avoid superluminal rotation, we assume that $|[\boldsymbol{\Omega},\boldsymbol{x}]|<c$ for all points of the system. This condition restricts the size of gas reservoir in the transverse to rotation axis direction. In what follows, we use a special form of the Gibbs distribution for a rotating system whose mass center is at rest,
\begin{equation}\label{f-pj-1}
    f_0(\varepsilon,\boldsymbol{j})=\frac{1}{z_0}\exp\bigg(-\frac{\varepsilon-(\boldsymbol{\Omega},\boldsymbol{j})}{kT}\bigg).
\end{equation}
The distribution function is not invariant with respect to the Poincare transformations because general transformations do not preserve the position of the mass center. However, the absence of the explicit Poincare symmetry in (\ref{f-pj-1}) do not cause the absence of the Lorentz invariance, because the original distribution function (\ref{f-pj}) has it. Hence, the hidden Lorentz symmetry is always present in the model.

For a classical ideal gas, the total energy and total angular momentum are given by the sum of corresponding quantities of individual particles, so the distribution function and partition function factorize into the product of one-particle contributions. Thus, the distribution function (\ref{f-pj-1}) can be applied to a single particle. In accordance with the previous computations, the particle energy $\varepsilon$ and total angular momentum $\boldsymbol{j}$ are represented in the following form:
\begin{equation}\label{ej-rep}
    \varepsilon=mc^2\gamma, \qquad \boldsymbol{j}=mc\gamma[\boldsymbol{x},\boldsymbol{\beta}]+\hbar\Sigma(\gamma\boldsymbol{c}-\frac{\gamma^2}{\gamma+1}\boldsymbol{\beta}(\boldsymbol{\beta},\boldsymbol{c}))\,.
\end{equation}
The solution uses the vector $\boldsymbol{c}$ (\ref{c-vec}), being parameterized by the angular variables $\alpha^\circ$, $\beta^\circ$. Both the quantities (\ref{ej-rep}) are considered as the functions of the phase-space variables $\boldsymbol{x}$, $\boldsymbol{p}$, $\alpha^\circ$, $\beta^\circ$. The invariant measure on the phase-space is determined by the formula (\ref{dg}). On substituting (\ref{ej-rep}) to (\ref{f-pj-1}), we obtain the one-particle distribution function for the rotating relativistic gas of spinning particles:
\begin{equation}\label{rho-distr-op}
f_0=\frac{1}{z_0}\exp\bigg(-\frac{mc^2\gamma}{kT}+\frac{mc\gamma(\boldsymbol{\Omega},[\boldsymbol{x},\boldsymbol{\beta}])}{kT}+\frac{\hbar\Sigma}{kT}\bigg(\boldsymbol{\Omega},\gamma\boldsymbol{c}-\frac{\gamma^2}{\gamma+1}\boldsymbol{\beta}(\boldsymbol{\beta},\boldsymbol{c})\bigg)\bigg)\,.  \end{equation}
The obtained one-particle distribution function determines a probability density to find a particle in a unit volume of phase-space in the vicinity of a point with the coordinates $\boldsymbol{x}$, $\boldsymbol{p}$, $\alpha^\circ$, $\beta^\circ$. The parameters of this function are angular velocity vector $\boldsymbol{\Omega}$ and thermodynamic temperature $T$. The (one-particle) partition function $z_0$ is determined from the normalizing condition,
\begin{equation}\label{z0-def}
    z_0=\int \exp\bigg(-\frac{mc^2\gamma}{kT}+\frac{mc\gamma(\boldsymbol{\Omega},[\boldsymbol{x},\boldsymbol{\beta}])}{kT}+\frac{\hbar\Sigma}{kT}\bigg(\boldsymbol{\Omega},\gamma\boldsymbol{c}-\frac{\gamma^2}{\gamma+1}\boldsymbol{\beta}(\boldsymbol{\beta},\boldsymbol{c})\bigg)\bigg)\mathrm{d}\Gamma\,.
\end{equation}
The integration is held over the invariant measure on the phase space. The round bracket with three enclosed vectors denote the mixed product of the vectors.  In what follows, we call (\ref{rho-distr-op}) the generalized Maxwell-Jüttner distribution for a rotating gas of spinning particles. Formula (\ref{rho-distr-op}) is a relativistic analogue of the generalized Maxwell-Boltzmann distribution, that has been derived in work \cite{BKN-2022}.

The distribution function (\ref{rho-distr-op}) can be rewritten in the rotating frame connected to the reservoir with gas. In the point with the coordinate $\boldsymbol{x}$, this frame moves with the velocity $[\boldsymbol{\Omega},\boldsymbol{x}]$ with respect to the inertial frame. If $c{t^\ast},{\boldsymbol{x}}^\ast$ denote the coordinates in rotating frame, the new and old variables are connected by the following Lorentz transformation: $ct=\upgamma(c{t}^\ast+({\boldsymbol{\upbeta}},{\boldsymbol{x}}^\ast))+o({t}^\ast)$, $\boldsymbol{x}=\upgamma({\boldsymbol{x}}^\ast+c{t}^\ast{\boldsymbol{\upbeta}})+o({t}^\ast)$. Making the Lorentz transformation from the inertial frame to the rotating frame, we rewrite the one-particle distribution function in the following form:
\begin{equation}\label{rho-distr-op-rf}
f_0=\frac{1}{z_0}\exp\bigg(-\frac{mc^2\gamma^\ast}{kT\upgamma}+
\frac{\hbar\Sigma\upgamma}{kT}\bigg(\boldsymbol{\Omega},\gamma^\ast{\boldsymbol{c}}^\ast-
\frac{(\gamma^\ast)^2}{\gamma^\ast+1}{\boldsymbol{\beta}}^\ast({\boldsymbol{\beta}}^\ast,{\boldsymbol{c}}^\ast)-(\gamma^\ast)^2[{\boldsymbol{\upbeta}},[{\boldsymbol{\beta}}^\ast,{\boldsymbol{c}}^\ast]]\bigg)\bigg)\,.  \end{equation}
Here, the parameters ${\gamma}^\ast$, ${\boldsymbol{\beta}}^\ast$ determine the relativistic factors of the particle in the rotating frame, and ${\boldsymbol{c}}^\ast$ determines the direction of spin in the rotating frame by the rule (\ref{c-vec}). The solution (\ref{rho-distr-op-rf})  uses the following transformation for the energy, momentum and spin angular momentum are expressed as follows: 
\begin{equation}\label{Lt-p}
    \varepsilon = mc^2\upgamma({\gamma}^\ast+({\boldsymbol{\upbeta}},{\boldsymbol{\beta}}^\ast))\,,\qquad \boldsymbol{p}=mc{\gamma}^\ast\bigg\{\upgamma({\boldsymbol{\beta}}^\ast+{\boldsymbol{\upbeta}}) - \frac{\upgamma^2}{\upgamma+1}({\boldsymbol{\beta}}^\ast{\upbeta}^2 - {\boldsymbol{\upbeta}}({\boldsymbol{\upbeta}},{\boldsymbol{\beta}}^\ast))\bigg\}\,,
\end{equation}
\begin{equation}\label{Lt-s}
    \boldsymbol{s} = \upgamma({\boldsymbol{s}}^\ast-{\gamma}^\ast[{\boldsymbol{\upbeta}},[\boldsymbol{{\beta}}^\ast,{\boldsymbol{s}}^\ast]])-\frac{{\boldsymbol{\upbeta}}\upgamma^2}{\upgamma+1}({\boldsymbol{s}}^\ast,{\boldsymbol{\upbeta}})\,.
\end{equation}
The relativistic factors ${\boldsymbol{\upbeta}}$, $\upgamma$ are determined by the rule
\begin{equation}\label{hat-beta}
    {\boldsymbol{\upbeta}}=\frac{1}{c}[\boldsymbol{\Omega},\boldsymbol{x}]\,,\qquad \upgamma=\frac{1}{(1-{\upbeta}^2)^{1/2}}\,.
\end{equation}
As the superluminal rotation is not allowed, the transformation (\ref{Lt-p}), (\ref{Lt-s}) is well-defined for all possible positions $\boldsymbol{x}$. The distribution function generalizes previously known Maxwell-Jüttner distribution in rotating frame \cite{MJ}. In the text below, we always determine the particle state in rotating frame. To avoid cumbersome notation, we systematically omit asterisk symbol in notations, e.g. $\gamma^\ast \equiv \gamma$.

The one-particle distribution function (\ref{rho-distr-op-rf}) has a special dependence on phase-space variables. The first contribution to the exponent argument determines the Maxwell-Jüttner distribution for spinless particle, while the spinning terms are expressed via the scalar product $(\boldsymbol{\xi},\boldsymbol{c})/kT$ of polarization vector $\boldsymbol{\xi}$ and unit vector $\boldsymbol{c}$. By construction,
\begin{equation}\label{xi}
    {\boldsymbol{\xi}}=    \hbar\Sigma\upgamma\bigg(\boldsymbol{\Omega}\gamma-
    \frac{\gamma^2}{\gamma+1}(\boldsymbol{\Omega},\boldsymbol{\beta})\boldsymbol{\beta}+\gamma^2[\boldsymbol{\beta},[\boldsymbol{\Omega},\boldsymbol{\upbeta}]]\bigg)\,.
\end{equation}
The special structure of angular distribution (\ref{rho-distr-op}) admits integration by directions of spin. This brings us to the one-particle distribution function by positions and momenta in rotating frame:
\begin{equation}\label{f-xp-in-rf}
    f_{xp}(\boldsymbol{x},\boldsymbol{p})=
    \frac{1}{z_0}\exp\bigg(-\frac{mc^2\gamma}{kT\upgamma}\bigg)\frac{\sinh\vartheta^\ast}{\vartheta^\ast}\,,
\end{equation}
where $\vartheta^\ast=\xi/kT$, where $\xi$ is the norm of polarization vector (\ref{xi}). The formula shows that the direction of spinning particle momenta becomes asymmetric due to interaction of spin and macroscopic rotation. As the function $\sinh\vartheta^\ast/\vartheta^\ast$ monotonously increases, the majority of linear momenta are directed in the plane orthogonal to the rotation axis. A similar expression for momentum-position distribution can be obtained from the inertial coordinate system, assuming $\upbeta = 0$.

The meaning of the obtained formulas (\ref{rho-distr-op}), (\ref{rho-distr-op-rf}), (\ref{f-xp-in-rf}) of the current work is that they determine one-particle distribution function by positions, momenta and directions of spin in rotating spinning particle gas. A similar problem of construction of position-momentum distribution has been addressed in a series of works \cite{Becc} by means of quantum formalism, the article  \cite{Florkowski} additionally uses quasi-classical treatment of spin. In our article, the spinning and translational degrees of freedom have been considered quasi-classically from the beginning. This allowed explicit construction of one-particle distribution function (\ref{rho-distr-op}) where the dependence on positions and momenta explicitly recovered. To our knowledge, the formulas (\ref{rho-distr-op}), (\ref{rho-distr-op-rf}) or its analogs have not been obtained previously withing quasi-classical formalism. In the context of physical interpretation of the formulas (\ref{rho-distr-op}), (\ref{rho-distr-op-rf}), it is important to mention that the spinning particle model (\ref{S-Ham}) admits consistent couplings with the general electromagnetic and/or gravitational field, with the corresponding field being taken at the particle position. This allows to consider the properties of rotating matter in external field by means of quasi-classical formalism in further studies. The use of quasi-classical formalism also solves the question of ambiguity of spinning particle positions, which is known for free spinning particle models, with the positions $\boldsymbol{x}$ being the fundamental coordinates of the particle. Moreover, the partition function $z_0$ determines the thermodynamic properties of the relativistic spinning particle gas. We address these problems below.

\section{Angular distribution function}

In the current section, we consider the distribution function $f_{\alpha\beta}$ by angular variables $\alpha^\circ$, $\beta^\circ$ in a fixed point space. There are two motivations for such problem. First, this function determines the angular distribution by the directions of particle's spin. Thus, a distribution function characterizes the behavior of spinning degree of freedom in the rotating systems. Second, the normalizing factor is appearing in this distribution will be used for computation of partition function $z_0$ (\ref{z0-def}) in the next section. 
As the angular distribution function varies from point-to-point, so we consider the particle position $\boldsymbol{x}$ as the parameter. 

It is convenient to solve the problem in the special coordinate system, where the angular velocity vector is co-directed with the third coordinate axis, $\boldsymbol{\Omega}=(0,0,\Omega)$. The particle position is supposed to be $\boldsymbol{x}=(x,0,0)$. This does not restrict the generality as the rotating system has a translational symmetry along the rotation axis, and the rotational symmetry around it. At the position $\boldsymbol{x}$, the rotational velocity has the form $\boldsymbol{\upbeta}=(0,\upbeta,0)$, $\upbeta=\Omega x/c$. The vector of momentum is expressed via a relativistic factor $\gamma$ and two angular variables $\xi^\circ$, $\zeta^\circ$, $\boldsymbol{p}=mc\sqrt{\gamma^2-1}(\cos\xi^\circ\sin\zeta^\circ,\sin\xi^\circ\sin\zeta^\circ,\cos\zeta^\circ) $. The angular distribution function $f_{\alpha\beta}$ in the accepted notation is determined by the following rule:
\begin{equation}\label{rho-distr-angles}
    f_{\alpha\beta}(\alpha^\circ,\beta^\circ) = \frac{1}{z_{\alpha\beta}}\int \exp\bigg(-\frac{mc^2\gamma}{kT\upgamma}+\frac{\hbar\Sigma\Omega}{kT}\upgamma\Phi\bigg)  \frac{\upgamma\mathrm{d}\boldsymbol{p}}{m^3c^3}\,,
\end{equation}
where $\Phi=\Phi_1+\Phi_2$ and the notation is used:
\begin{equation}\begin{array}{c}\displaystyle
    \Phi_1 = \gamma \cos\beta^\circ - (\gamma-1)\cos\zeta^\circ(\sin\beta^\circ \sin\zeta^\circ \cos(\alpha^\circ - \xi^\circ) + \cos\beta^\circ \cos\zeta^\circ)\,,\\[5mm]\displaystyle
    \Phi_2 = \upbeta\gamma\sqrt{\gamma^2-1}(\cos\beta^\circ \sin\zeta^\circ \sin\xi^\circ - \sin\alpha^\circ \sin\beta^\circ \cos\zeta^\circ)\,.
\end{array}   
\end{equation}
Here, $\mathrm{d} \boldsymbol{p} = m^3c^3 \sqrt{\gamma^2-1}\gamma \mathrm{d} \gamma \sin\zeta^\circ \mathrm{d}\xi^\circ\mathrm{d}\zeta^\circ$ is the volume element in the momenta space, expressed in the rotating frame. The factor $\upgamma$ in (\ref{rho-distr-angles}) appears as the result of Lorentz transformation.  
The partition function $z_{\alpha\beta}$ form the normalizing condition,
\begin{equation}\label{z_ab-rule}
    \frac{1}{4\pi}\int f_{\alpha\beta}\sin\beta^\circ \mathrm{d}\alpha^\circ \mathrm{d}\beta^\circ=1\,.
\end{equation}
In the accepted normalization condition, the uniform angular distribution corresponds to $f_{\alpha\beta}=1$.

The distribution function (\ref{rho-distr-angles}) involves two dimensionless parameters, $\theta=kT/mc^2$ and $\vartheta=\hbar\Sigma\Omega/kT$. The quantity $\theta$ serves dimensionless temperature, and $\vartheta$ characterizes the magnitude of chiral effects. The multiplication of the parameters by $\upgamma$ gives another pair of dimensionless quantities $\uptheta=\theta\upgamma$, $\upvartheta=\vartheta\upgamma$. The article \cite{BKN-2022} tells us that the value of the parameter $\vartheta$ (and, hence, $\upvartheta$) is very small, being of order $10^{-10}$ at the degeneration temperature for non-relativistic gas. As the relativistic effects appear in the hot gases, we expand the spinning part of the distribution (\ref{rho-distr-angles}) into the power series in $\upvartheta$ only up to the second order in spin. By direct integration of the obtained expression by momenta, we get the following form of angular distribution function:
\begin{equation}\label{op-ang-final}
    f_{\alpha\beta}(\alpha^\circ,\beta^\circ)=1+4\pi\upgamma\uptheta\frac{\mathcal{A}_2(\alpha^\circ,\beta^\circ;
    \uptheta,\upvartheta)\mathrm{K}_2(\uptheta)+
    \mathcal{A}_3(\alpha^\circ,\beta^\circ;\uptheta,\upvartheta)\mathrm{K}_3(\uptheta)}{z_{\alpha\beta}}\,,
\end{equation}   
where $\mathrm{K}_a(\uptheta)=K_a(1/\uptheta)$, $a=2,3$ are modified Bessel's functions of inverse argument, and the notation is used
\begin{equation}\label{A2}\begin{array}{c}\displaystyle
\mathcal{A}_2=\frac{1-2\uptheta}{3} \upvartheta\cos\beta^\circ  +\upvartheta^2\bigg(\frac{3 -2\uptheta}{30}(3\cos^2\beta^\circ
-1) +\frac{5}{6}\uptheta^2\upbeta^2(1-3\cos^2\alpha^\circ \sin^2\beta^\circ)\bigg);
\end{array}\end{equation}
\begin{equation}\label{A3}\begin{array}{c}\displaystyle
\mathcal{A}_3=\frac{2}{3}\upvartheta\cos\beta^\circ +\upvartheta^2\bigg( 
\frac{2 + 7\uptheta}{30}(3\cos^2\beta^\circ-1)+ \frac{\uptheta + 30\uptheta^3}{6}\upbeta^2(1-3\cos^2\alpha^\circ \sin^2\beta^\circ)\bigg).
\end{array}\end{equation}
The partition function $z_{\alpha\beta}$ reads
\begin{equation}\label{z0-exp}
    z_{\alpha\beta}=4\pi\upgamma\bigg\{\uptheta\bigg(1+\upvartheta^2\frac{1+10\uptheta^2\upbeta^2}{6}\bigg)\mathrm{K}_2(\uptheta)+\uptheta^2\upvartheta^2\frac{1+\upbeta^2(1+30\uptheta^2)}{3}\mathrm{K}_3(\uptheta)\bigg\}\,.
\end{equation}
The formula (\ref{op-ang-final}), (\ref{A2}), (\ref{A3}) with the normalizing factor (\ref{z0-exp}) determines the angular distribution function with account of spinning corrections up to second order. The form of function $f_{\alpha\beta}$ shows that the angular distribution function is anisotropic due to chiral effects. In the leading order in spin (entering into the parameter $\upvartheta$, we have dependence on the angle $\beta^\circ$, that corresponds to the polarization of spinning degree of freedom. As the distribution function $f_{\alpha\beta}$ decreases with growth of $\beta^\circ$, there is dominating orientation of spins directed along the rotation axis. In the second order in spin, we have the dependence on two angular variables $\alpha^\circ$, $\beta^\circ$, so the distribution of spins in the orthogonal to rotation direction becomes asymmetric. This effect is caused by the use of rotating coordinate system. For $\upbeta=0$, the distribution function (\ref{op-ang-final}), (\ref{A2}), (\ref{A3}) describes the one-particle distribution function at the rotation axis in the inertial frame. This distribution function depends on a single angle $\beta^\circ$. 

In order to visualize the angular distribution, it is convenient to introduce the distribution functions $f_\alpha$, $f_\beta$ by a single variable, which introduce by the rule:
\begin{equation}
    f_\alpha(\alpha^\circ)=\frac{1}{2}\int f_{\alpha\beta}\,\sin\beta^\circ \mathrm{d}\beta^\circ\,,\qquad f_{\beta}(\beta^\circ)=\frac{1}{2\pi}\int f_{\alpha\beta}\mathrm{d}\alpha^\circ\,.
\end{equation}
By constriction, the first spinning correction to the distribution function $f_\beta$ is proportional to first power of $\upvartheta$, while for $f_\alpha$ the correction starts from $\upvartheta^2$. In the leading order of $\upvartheta$, we have the following estimates:
\begin{equation}\label{f-alpha}
    f_\alpha =1-\frac{5\uptheta \mathrm{K}_2(\uptheta)+(1+30\uptheta^2)\mathrm{K}_3(\uptheta)}{6\mathrm{K}_2(\uptheta)}\uptheta\upbeta^2\upvartheta^2\cos(2\alpha^\circ)+o(\upvartheta^2)\,.
\end{equation}
\begin{equation}\label{f-beta}
    f_{\beta}=1+\frac{\displaystyle(1-2\uptheta)\mathrm{K}_2(\uptheta)+2 \mathrm{K}_3(\uptheta)}{\displaystyle  3 \mathrm{K}_2(\uptheta)}\upvartheta\cos\beta^\circ+o(\upvartheta)\,.
\end{equation}
Figure 1 shows the polar plot of function $f_{\alpha}$ (\ref{f-alpha}) for selected values of the parameters. As is seen from the plot, the shape of the figure deforms with the increase of the parameter $\upbeta$ in accordance with equation (\ref{hat-beta}). The most probable orientation of spin at the current plane is determined by the direction to/from rotation axis. The orientation of spin along the velocity vector $\upbeta$ is the least probable. 
Figure 2 shows the plot of function $f_{\alpha}$ (\ref{f-beta}) for selected values of the parameters. As is seen from the plot, the shape of the angular distribution function becomes elongated with the increase of the parameter $\upvartheta$. The most probable orientation of spin is determined by the angular velocity vector $\boldsymbol{\Omega}$.

\begin{figure}\label{fa}
\includegraphics[scale=0.9]{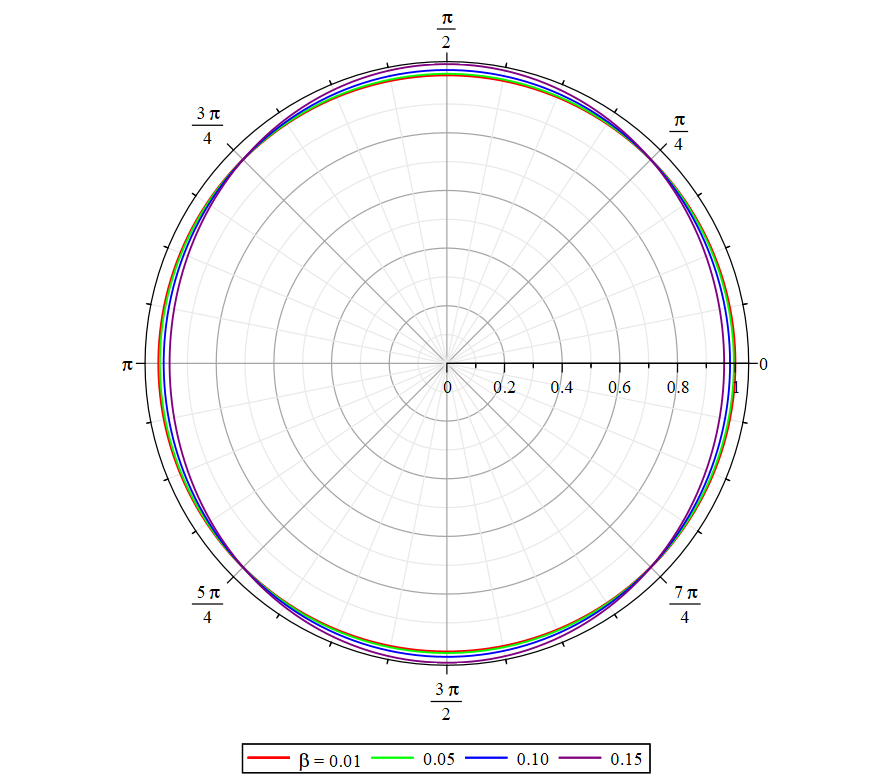}
\centering
\caption{Angular distribution function $f_\alpha$ for $\uptheta=1$, $\upvartheta=0.3$. }
\end{figure}

\begin{figure}\label{fb}
\includegraphics[scale=0.9]{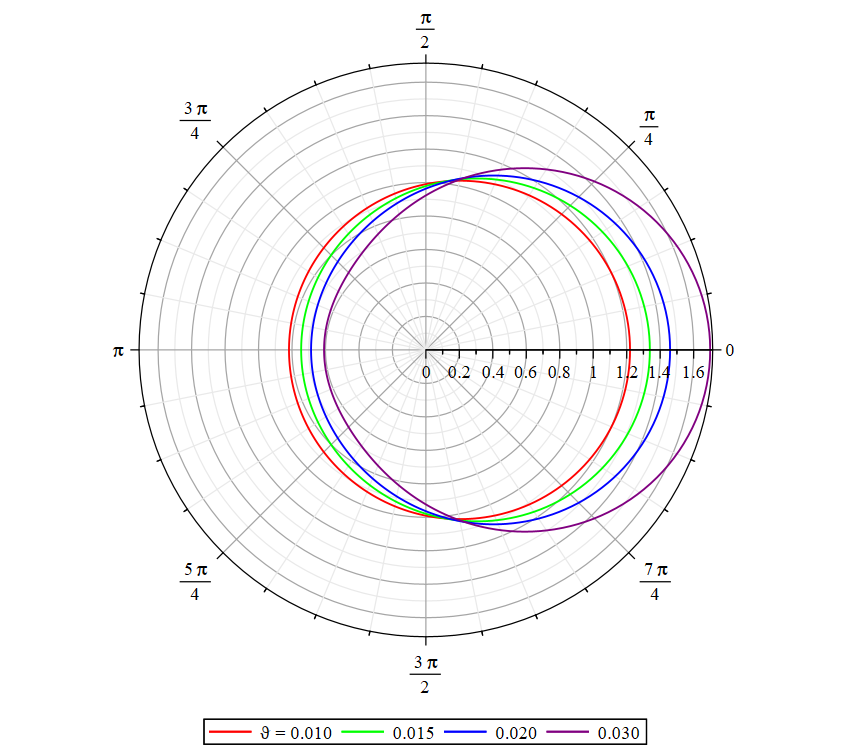}
\centering
\caption{Angular distribution function $f_\beta$ for $\upbeta=0.001$, $\uptheta=10$. }
\end{figure}

\section{Thermodynamics of rotating gas}

In this section, we consider the thermodynamics of a slowly rotating relativistic ideal gas of spinning particles. For simplicity reasons, we assume that the gas reservoir have the shape of cylinder with the height $h$ and the radius $R$. The slow rotation of the cylinder suggests that the linear velocity of reservoir rotation is much smaller than the speed of light, i.e. $\beta_m=\Omega R/c\ll1$. We also assume that the magnitude of chiral effects is small, so $\vartheta\ll1$. The partition function (\ref{z0-def}) involves the squares of small parameters, so the leading rotational correction to the thermodynamic potential is determined by the terms linear in $\beta_{m}^2$ and $\vartheta^2$. Our consideration mostly follows the lines of work \cite{BKN-2022}.

At first, we need to compute the partition function $z_0$ (\ref{z0-def}) with account of the known form of gas reservoir. As the integration over the momenta and angular variables has been already computed in the previous section, it is convenient to express the (one-particle) partition function $z_0$ (\ref{z0-def}) via the angular partition function $z_{\alpha\beta}$ (\ref{z0-exp})\,. The partition function of classical ideal gas consisting of $N$ particles is determined by the rule:
\begin{equation}\label{z-def}
    z=\frac{1}{N!}(z_0)^N\,.
\end{equation}
Here, the combinatorial factor $N!$ accounts to the number of permutations of $N$ particles. We assume that $N$ is a very big number, so the limit $N\to\infty$ is applied below. The partition function (\ref{z-def}) determines the thermodynamic potential $\Phi$ of the gas by the rule,
\begin{equation}\label{p-z}
    \Phi=-kT\ln z\,.
\end{equation}
The natural variables of the thermodynamic potential are temperature $T$, angular velocity $\Omega$, and the particle number. The thermodynamic potential serves as the analog of Gibbs free energy for a rotating system. The following Legendre transformation connects the thermodynamic potential with the internal energy: $\Phi=U-TS-\Omega j$, with $S,j$ respectively being entropy and total angular momentum of the gas. The differential of $\Phi$ reads $\mathrm{d}\Phi=-S\mathrm{d}T-j\mathrm{d}\Omega+\mu \mathrm{d}N$. with $\mu$ being the chemical potential per particle. The extensive properties of thermodynamic potential also imply $\Phi=\mu N$.

Using the decomposition, we find the partition function for relativistic gas in slowly rotating cylinder,
\begin{equation}
    z_0=\frac{4\pi m^3c^3(2s+1) V}{(2\pi\hbar)^3}\theta \mathrm{K}_2(\theta)\bigg(1+\beta_m^2\frac{\mathrm{K}_3(\theta)-2\theta\mathrm{K}_2(\theta)}{4\theta\mathrm{K}_2(\theta)}+\vartheta^2\frac{\mathrm{K}_2(\theta)+2\theta\mathrm{K}_3(\theta)}{6\mathrm{K}_2(\theta)}+\ldots \bigg)\,.
\end{equation}
The parameters of the expression are dimensionless temperature $\theta$, and parameters $\beta_m$, $\vartheta$ that involve the angular velocity. The leading term of the expression determines the thermodynamic potential of non-rotating relativistic gas, and two other terms account rotational and spinning corrections. Applying the formula (\ref{p-z}), we find the thermodynamic potential of the gas with account of corrections,  
\begin{equation}
    \Phi=-kTN\bigg\{\ln \frac{4\pi m^3c^3(2s+1) V}{N(2\pi\hbar)^3}\theta \mathrm{K}_2(\theta)+\beta_m^2\frac{\mathrm{K}_3(\theta)-2\theta\mathrm{K}_2(\theta)}{4\theta\mathrm{K}_2(\theta)}+\vartheta^2\frac{\mathrm{K}_2(\theta)+2\theta\mathrm{K}_3(\theta)}{6\mathrm{K}_2(\theta)}+1\bigg\}\,.
\end{equation}
Differentiating this expression with respect to natural variables $T$, $\Omega$, and $N$, we obtain the thermodynamic parameters of the system. As the ration $\Phi/N$ determine the chemical potential, only the derivative with respect to $T$ and $\Omega$ are relevant. Differentiating by $\Omega$ (with minus sign) leads us to the total angular momentum of the system:
\begin{equation}
    j=N\bigg\{\frac{m\Omega R^2}{2}\frac{\mathrm{K}_3(\theta)-2\theta\mathrm{K}_2(\theta)}{\mathrm{K}_2(\theta)}+\frac{\hbar\Sigma\vartheta}{3}\frac{\mathrm{K}_2(\theta)+2\theta\mathrm{K}_3(\theta)}{\mathrm{K}_2(\theta)}\bigg\}\,.
\end{equation}
Both terms in the obtained expression have clear physical interpretation. The $mR^2\Omega/2$ contribution have a sense of orbital angular momentum of spinless relativistic gas. It is interesting to note that the momentum of inertia of the gas depends on temperature. The average spin projection onto the rotation axis can be found as the spin angular momentum per particle:
\begin{equation}\label{s-avg}
    \langle s\rangle=\frac{\hbar\Sigma\vartheta}{3}\frac{\mathrm{K}_2(\theta)+2\theta\mathrm{K}_3(\theta)}{\mathrm{K}_2(\theta)}.    
\end{equation}
In the motion of low temperatures, this formula gives the non-relativistic polarization expression of the work \cite{BKN-2022}, $\langle s\rangle=\hbar\Sigma\vartheta/3$.

The thermodynamic potential significantly simplifies in the ultra-relativistic case: $\theta\to\infty$. Using the modified Bessel's function estimate for small values of argument, $\mathrm{K}_n\approx\Gamma(n)(2\theta)^n/2$, we obtain
\begin{equation}
    \Phi=-kTN\bigg\{\ln \frac{8\pi m^3c^3(2s+1)V}{N(2\pi\hbar)^3}\theta^3+\frac{1}{2}\beta_m^2+\frac{4}{3}\theta^2\vartheta^2+1\bigg\}\,.
\end{equation}
For a narrow cylinder, $\beta_m=0$, this expression has a form of thermodynamic potential of ideal gas with constant heat capacity $C_V=3N$ with the special value of chemical constant, being determined by the particle mass and spin. The entropy of relativistic gas reads:
\begin{equation}
    S=kN\bigg\{\ln \frac{8\pi m^3c^3(2s+1)V}{N(2\pi\hbar)^3}\theta^3+\frac{1}{2}\beta_m^2+\frac{4}{3}\theta^2\vartheta^2+4\bigg\}\,.
\end{equation}
It is interesting to mention that the rotational term determines temperature-independent additive contribution. This means that the rotation changes the entropy of the gas, respectively to temperature. As the chiral corrections in the non-relativistic gas decrease with the growth of temperature, the class of ultra-hot gases can be considered as a good system for observation of chiral effects.  Differentiating by $\Omega$ (with minus sign) leads us to the total angular momentum of the ultra-relativistc gas:
\begin{equation}
    j=N\bigg\{m\Omega R^2\theta+\frac{8}{3}\hbar\Sigma\vartheta\theta^2\bigg\}\,.
\end{equation}  
It is interesting to mention that the average angular momentum linearly grows with the temperature in the ultra-relativistic limit. The average value of projection of spin angular momentum onto the rotation axis reads:
\begin{equation}
   \langle s\rangle =\frac{8}{3}\hbar\Sigma\vartheta\theta^2\,.
\end{equation} 
As is seen, average projection of spin angular momentum also linearly increases with the temperature. This indicates an increase of significance of chiral effects in the relativistic gases.

Now, let us consider the specific rotational polarizability of spinning degree of freedom,
\begin{equation}
    \chi\equiv\frac{\partial\langle s\rangle}{\partial \Omega}\bigg|_{\Omega=0}=\frac{\hbar^2\Sigma^2}{3mc^2}\frac{\mathrm{K}_2(\theta)+2\theta\mathrm{K}_3(\theta)}{\theta\mathrm{K}_2(\theta)}\,.
\end{equation}
\begin{figure}\label{fb}
\includegraphics[scale=0.8]{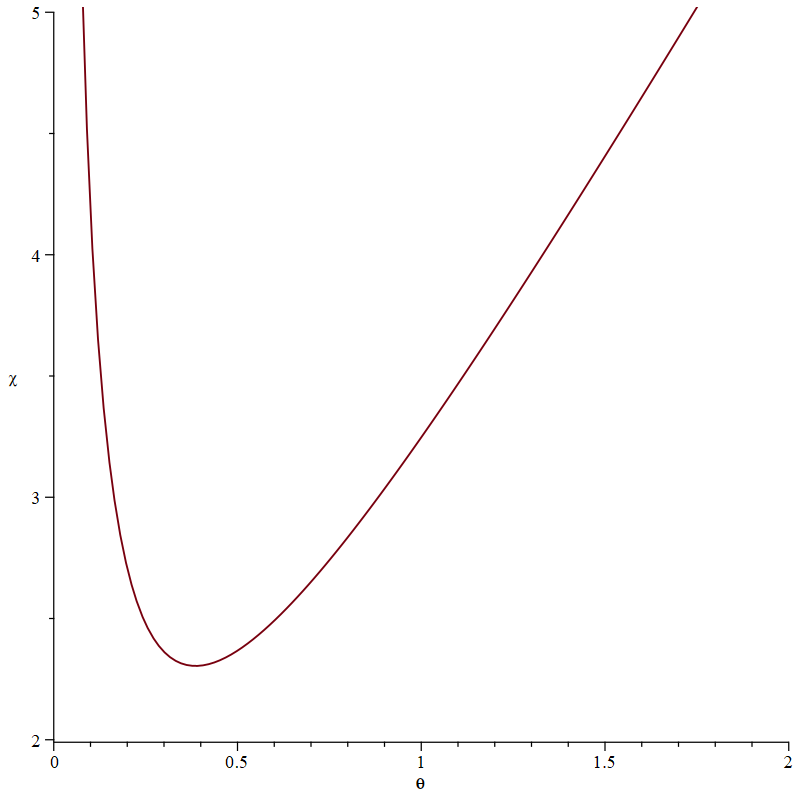}
\centering
\caption{Specific polarizability of relativistic gas $\chi(\theta)$}
\end{figure}
By construction, this quantity determines the change of average spin angular momentum projection on the rotation axis per unit change of angular velocity. Figure 3 shows a plot of specific polarizability of relativistic gas $\chi mc^2/\hbar^2\Sigma^2$ depending on the dimensionless temperature $\theta$. As is seen, the polarizability decreases for low temperatures. This part of the plot is described by the non-relativistic formula of work \cite{BKN-2022}. For high values of $\theta$, the polarizability increases. At the intermediate value of temperature, we have minimum at $\theta_{m}=0.3859...$ with the extreme value $\chi_m = 2.3044...\hbar^2\Sigma^2/mc^2$. To our knowledge, the presence of polarizability minimum at intermediate temperatures of relativistic spinning rotating gas has not been observed in the previous studies. 

\section{Comparison with results of Ref. \cite{Becc}}

In the article \cite{Becc}, the quantum version of rotational grand-canonical distribution function is applied to the rotating gas of spinning particles. It has been shown that the distribution function admits re-summation such that the spinning and translational contributions decouple. This allowed explicit construction of a distribution function by positions and momenta generalizing the usual Maxwell-Jüttner distribution for rotating relativistic ideal gas. Below, we demonstrate that the classical distribution function of the work \cite{Becc} agrees with the quasi-classical formula (\ref{rho-distr-op}) in the infinite spin limit. The crucial observation here is that the classical position $\boldsymbol{x}$ of the spinning particle has the nonzero Dirac brackets due to the presence of constraints. The set of canonically conjugated variables has been derived in \cite{DP-2014}, and these variables correspond to the momenta and position in the Foldy–Wouthuysen representation \cite{DP-2016}. 

Following the work \cite{DP-2014}, we introduce the canonical variables $\boldsymbol{x}'$, $\boldsymbol{p}'$, $(\alpha^\circ)'$, $(\beta^\circ)'$ on the phase-space of a spinning particle by the following rule:
\begin{equation}\label{prime-var}
   \boldsymbol{x}=\boldsymbol{x}'-\frac{\gamma'}{\gamma'+1}\frac{[\boldsymbol{s}', \boldsymbol{\beta}']}{mc}\,,\qquad \boldsymbol{p}=\boldsymbol{p}'\,,\qquad \alpha^\circ=(\alpha^\circ)',\qquad \beta^\circ=(\beta^\circ)'.
\end{equation}
Here, the spin angular momentum vector $\boldsymbol{s}'=\hbar\Sigma\boldsymbol{c}'$ is determined by the usual rule (\ref{c-vec}) with $\alpha^\circ$, $\beta^\circ$ being replaced by primed quantities $(\alpha^\circ)'$, $(\beta^\circ)'$, and $\boldsymbol{p}'=mc\widetilde\gamma'\boldsymbol{\beta}'$. We note that the transformation is identical for non-relativistic particle with $\beta=0$. In the new variables, the energy and total angular momentum take the following form:
\begin{equation}\label{j-tilde}
    \varepsilon'=mc^2\gamma'\,,\qquad \boldsymbol{j}=[\boldsymbol{x}',\boldsymbol{p}']+\hbar\Sigma\boldsymbol{c}'\,.
\end{equation}
By construction, the vector $\boldsymbol{c}'$ determines the direction of particle spin in the canonical coordinates. The change of the variables (\ref{prime-var}) is unimodular, so the phase-space element has the form (\ref{dg}) with the original variables being replaced by the primed ones. On substituting the expression for the energy and total angular momentum (\ref{j-tilde}) into the distribution function, the following result is obtained for the phase-space density:
\begin{equation}\label{fxpab-canonical}
    f(\boldsymbol{x}', \boldsymbol{p}', (\alpha^\circ)', (\beta^\circ)')=\frac{1}{z_0}\exp\bigg(-\frac{mc^2\gamma'-(\boldsymbol{\Omega},[\boldsymbol{x}',\boldsymbol{p}']+\hbar\Sigma\boldsymbol{c}'((\alpha^\circ)', (\widetilde{\beta}^\circ)'))}{kT}\bigg)\,.
\end{equation}
The expression allows integration by the angular variables $(\alpha^\circ)', (\beta^\circ)'$ along the lines of Section 3. For the distribution function in canonical positions and momenta, $\boldsymbol{x}'$ and $\boldsymbol{p}'$, we get the result of the article \cite{Becc},
\begin{equation}\label{fxp-canonical}        f(\boldsymbol{x}',\boldsymbol{p}')=\frac{1}{z_0}\exp\bigg(-\frac{mc^2\gamma'-(\boldsymbol{\Omega},[\boldsymbol{x}',\boldsymbol{p}'])}{kT}\bigg)\frac{\sinh\vartheta}{\vartheta}\,.
\end{equation}
The formula agrees with the result of quantum computations of \cite{Becc} in the limit $s\to\infty$. This restores a relationship of quantum and classical formalism.

It is important to note that the construction of distribution (\ref{fxp-canonical}) essentially uses the absence of interaction with external fields. In the model with coupling, the interaction is determined in the unprimed coordinates $\boldsymbol{x}$, while in the Foldy–Wouthuysen phase-space variables (\ref{prime-var}) it becomes non-local. In the context of couplings with electromagnetic field, the discussion on the subject can be found in \cite{DP-2016}. The non-locality of interaction in the Foldy–Wouthuysen variables also imply that the function (\ref{fxp-canonical}) represents somewhat averaged distribution by spins in the vicinity of the point with the coordinate $\boldsymbol{x'}$ with the size of order $\hbar\Sigma\beta/mc$. The distribution (\ref{fxp-canonical}) cannot be used for computation of partition function $z_0$ (\ref{z-def}) as the quantity $\boldsymbol{x}'$ does not correspond to real position of the particle. In particular, the coordinate $\boldsymbol{x}'$ cannot be restricted by the shape of the gas reservoir, as the interaction becomes non-local in the setting. If the positions $\boldsymbol{x}$ are replaced $\boldsymbol{x'}$, the thermodynamic parameters of gas gets proportional to $\hbar^2$ corrections. These corrections become significant in the ultra-relativistic limit. The polarization of spinning degree of freedom increase in the hot gas, even though in the Foldy–Wouthuysen variables (\ref{prime-var}) spinning corrections monotonously decrease. The distribution function (\ref{fxpab-canonical}) determines the distribution by positions in a point of physical space only in the non-relativistic limit when the coordinates $\boldsymbol{x}$ and $\boldsymbol{x}'$ are equal. In the last case, the resulting distribution agrees with the non-relativistic formula of work \cite{BKN-2022}. 

\section{Conclusion}

In the current article, we have considered the statistical physics and thermodynamics of rotating ideal gas of classical relativistic spinning particles with nonzero mass. Using the canonical Gibbs ensemble with constant values of the temperature $T$ and angular velocity $\boldsymbol{\Omega}$, we found the one-particle distribution function by positions, momenta and spin variables, that generalizes the well-known Maxwell-Jüttner distribution. We have observed that the rotation causes the polarization of spinning degree of freedom, with the majority of spins being directed along the angular velocity vector. That demonstrates the presence of chiral phenomena in the model. By finding the partition function, we obtained the thermodynamic potential of the gas in the (slowly) rotating cylinder, and computed the macroscopic parameters such as entropy, angular momentum and chemical potential. All these quantities demonstrate chiral corrections, being proportional to angular velocity and spin. The final formulas significantly simplify for ultra-hot gas, and they describe an ideal gas with constant heat capacity. We have also observed that the polarizability of spin has the minimum at the intermediate values of temperatures of rest energy order. For lower and higher temperatures, the spin polarization increases. The result seems to be interesting because it suggests high value of chiral effects in very hot systems. This effect is relativistic because the non-relativistic distribution function suggests monotonous decrease of chiral effects at high temperatures. We have also demonstrated that our results agree with the previous studies of classical non-relativistic gas \cite{BKN-2022}, and quantum and quasi-classical computations \cite{Becc}, \cite{second reviewer ed1}. The obtained results demonstrate that the quasi-classical theory of spin can be applied to study of chiral properties of relativistic gases. As the possible application of the theory, we can mention the statistical mechanics of high-spin rotating gases in external electromagnetic and/or gravitational field.

\section{Acknowledgements}
The authors thank Yu.V. Brezhnev, P.O. Kazinski, N. Makhaldiani, A.A. Sharapov, and anonymous referees for valuable comments of this work. We express our special gratitude to I.V. Gorbunov for discussions concerning the classical description of spin. The work was supported by the RSF project 21-71-10066.

\renewcommand\refname{Bibliography}

\end{document}